\documentstyle[12pt]{article}
\begin{document}

\begin{center} The Effects of Deformation on Isovector \\
Electromagnetic and Weak Transition Strengths \medskip \\ L. Zamick
and N. Auerbach$^{*}$ \\ Department of Physics and Astronomy \\
Rutgers University \\ Piscataway, NJ  08855
\end{center}

\begin{abstract}
The summed strength for transitions from the ground state of $^{12}C$
via the operators $\vec{s}t, \vec{\ell}t, rY^{\prime}t,
r[Y^{\prime}s]^{\lambda}t$ and $r[Y^{\prime}\ell]^{\lambda}t$ are
calculated using the $\Delta N = 0$ rotational model.  If we choose
the z component of the isospin operator $t_{z}$, the above operators are
relevant to electromagnetic transitions; if we choose $t_{+}$ they are
relevant to weak transitions such as neutrino capture.  In going from
the spherical limit to the asymptotic (oblate) limit the strength for
the operator $\vec{s} t$ decreases steadily to zero; the strength for
the operator $\vec{\ell}\tau$ (scissors mode) increases by a factor of
three.  For the last three operators - isovector dipole, spin dipole
and orbital dipole (including the twist mode) it is shown that the
summed strength is independant of deformation.  The main difference in
the behavior is that for the first two operators we have in-shell
transitions whereas for the last three operators the transitions are
out of shell.
\end{abstract}
\vspace{2.5in}

\noindent $^{*}$Permanent address, School of Physics, Tel-Aviv
University, Tel Aviv, Israel.

\newpage
\noindent 1)  Introduction
\bigskip

The present generation of neutrino-oscillation experiments [1] require
the knowledge of neturino-nucleus reaction cross-sections of both
exclusive and inclusive processes.  In particular, neutrino reactions 
on the $^{12}C$ nucleus have been very instrumental in studying
neutrino oscillations in the LSND [2] and KARMEN [3] experiments.

Neutrino flux averaged $^{12}C$($\nu_{\mu},\mu^{-}$) and
$^{12}C(\nu_{e},e^{-})$ inclusive cross section as well as the
exclusive $^{12}C(\nu_{\mu}, \mu^{-})^{12}N_{g.s.}$ and
$^{12}C(\nu_{e}, e^{-})^{12}N_{g.s.}$ were determined in the
above experiments.  Theoretical and experimental attention has been
given to the $^{12}C(\nu_{\mu}, \mu^{-})$ measurement performed at
LSND with a flux of neutrinos obtained from the decay in flight (DIF) of
pions.  The experimental flux averaged cross-section to the $^{12}N$
ground state is [4].  

\begin{equation}
\sigma_{exc} = (0.66 \pm 0.1 \pm 0.1) \times 10^{-42} cm^{2}
\end{equation}
while the flux averaged total inclusive cross-section (which includes
the transition to $^{12}N_{g.s.}$ is now put at [4],
\begin{equation}
\sigma_{inc} = (12.2 \pm 0.3 + 1.8) \times 10^{-42} cm^{2}
\end{equation}
When the $^{12}C$ ground state is assumed to form a $p^{8}_{3/2}$
closed shell the Hartree-Fock (HF) and the subsequent Random Phase
Approximation (RPA) calculation give an exclusive cross-section which
is a factor 5-7 [5,6] too large compared to the LSND result.  The
inclusive cross-section in the RPA turns out to be $(20-22) \times
10^{-42} cm^{2}$ [5,6] almost a factor of two larger than the LSND
result.  However, it is well known for a long time that the
$^{12}C$ ground state is not well described by a pure $p^{8}_{3/2}$
configuration.  The Gamow-Teller transition to the $J = 1^{+}$ g.s. of
$^{12}N$ as well as the M1 transition in $^{12}C$ are too large by
factors of about 5 when a pure $p^{8}_{3/2}$ configuration is used for
the $^{12}C$ g.s.  Configuration mixing and the use of intermediate
coupling wave functions reduces these transitions.  In the early
theoretical work [5] the authors have accounted for the fact that the
$p^{8}_{3/2}$ g.s. is not a good starting point by simply dividing the
calculated exclusive $(\nu_{\mu,}\mu^{-})$ cross section by
approximately a factor of five.  They have however not altered the
inclusive cross-section to the excited states, leaving unchanged the
RPA result, 
which was calculated starting from a pure $p^{8}_{3/2}$ g.s.  In
ref. [6] the reduction of the exclusive cross-section was achieved by
using a pairing model approximation.  To account for the configuration
mixing in the g.s. of $^{12}C$ a Bogoliubov-Valatin approximation was
applied and partial occuppation numbers for the $p_{3/2}$ and
$p_{1/2}$ orbits were assigned, $(v_{p_{3/2}} = 1, v_{p_{1/2}} = 0)$.
This leads to a large reduction in the exclusive cross section which
is determined completely by the Gamow-Teller strength of the $J=1^{+}$
g.s. in $^{12}N$.  In a very approximate way the effects of pairing on
the inclusive cross-sections were also estimated.  These estimates,
which included also the energy shift due to pairing, led to a
20-25\% reduction.

In this work we wish to study the effects of deformation on the
transitions from the ground state of $^{12}C$.  Indeed $^{12}C$ is
generally regarded as being strongly deformed oblate nucleus and a
deformation parameter $\beta \sim -0.6$ is often quoted.  One cannot
simulate deformation effects by doing an RPA calculation on a
\underline{spherical} ground state.  In nuclei where pairing
correlations are important, both deformation and pairing should be
taken into account.  It is not clear how important pairing is for the
light nuclei.  

We feel that not too much is known on the effects of deformation on
\underline{isovector} transitions -- the main focus has been on the
isoscalar enhancement of the B(E2) $0_{1} \rightarrow 2$, in ground
state rotational bands.  We therefore shall focus our attention on
simple isovector operators which are relevant to both electromagnetic
and weak transitions, and we shall consider the simplest problem of
summed strengths to states ($\lambda , K$) where $\lambda$ is the
final total angular momentum and K the projection of total angular
momentum on the symmetry axis in the intrinsic state of a deformed
rotating nucleus.
\bigskip

\noindent 2)  $^{12}C$ in the Spherical and Asymptotic Limits
\medskip

In the spherical limit $^{12}C$ consists of 4 nucleons in the
0s$_{\frac{1}{2}}$ shell and eight in the $p_{\frac{3}{2}}$ shell.  An
elastic dipole transition could take place either by exciting a
nucleus from $op_{\frac{3}{2}}$ to the 1s-0d shell or by exciting a
0s$_{\frac{1}{2}}$ nucleon to $p_{3/2}$.  An M1 transition would consist
of exciting a $p_{3/2}$ nucleon to the $p_{1/2}$ shell.

It was shown very early on e.g. by Cohen and Kurath [7] that this limit is
a very bad approximation.  There is considerable configuration mixing.
In the spherical limit the $J = 1^{+}$ state wuld consist of seven
$p_{3/2}$ and one $p_{1/2}$ nucleon.  However, in shell model
calculations one finds that the number of $p_{1/2}$ nucleons in the $J
= 1^{+}_{1}$ state is almost the same as in the ground state.

We next consider the asymptotic oblate limit.  It is generally
believed that the ground state band in $^{12}C$ is oblate with $\beta
\approx -0.6$.  Indeed in the deformed oscillator model with the
Mottelson conditions one obtains such a result.  We may therefore
expect that this limit is closer to reality than the spherical limit.

In the asymptotic oblate limit we have the following single-particle
levels and wavefunctions.

\begin{tabbing}
xxxx \= xxxxxxx \= xxxxxxxx \= xxxxxxxxxx \= xxxxxxxxxxx \=
xxxxxxxxxxx \kill

1) \> N=0 \> $Y_{00}\uparrow$ \> $(Y_{00}\downarrow)$ \>
$(K=\frac{1}{2})$ \> occupied \\
2) \> N=1 \> $rY_{11}\uparrow$ \> $(Y_{1-1}\downarrow)$ \> $(K =
\frac{3}{2})$ \> occupied \\
3) \> N=1 \> $rY_{11}\downarrow$ \> $(Y_{1-1}\uparrow)$ \> $K =
\frac{1}{2}$ \> occupied \\
4) \> N=1 \> $rY_{10}\uparrow$ \> $(Y_{10}\downarrow)$ \> $K =
\frac{1}{2}$ \> empty \\
\end{tabbing}

We can see immediately that the matrix element of the spin
operators s 
or st from the $J=0^{+}$ ground state to an excited $J=1^{+}$ state in
$^{12}C$ will vanish.  The spin operator cannot change $Y_{11}$ into
$Y_{10}$.  The result is more general--in the SU(3) limit the spin 
M1's will vanish. 
\bigskip
\medskip

\noindent 3)  The M1 Transition in $^{12}C$ 
\medskip

We quote the results and use the conventions of reduced motion elements of
Bohr and Mottelson [8].
\begin{equation}
\vec{M}1 = \sqrt{\frac{3}{4}\pi} \vec{\mu} 
\end{equation}
\begin{eqnarray}
<j_{2} & = & \ell +
\frac{1}{2}||M1|| j_{1} = \ell - \frac{1}{2}> \nonumber \\ 
& = &  -\sqrt{\frac{3}{4\pi}}
(g_{s}-g_{\ell}) \sqrt{2j_{1}+1}  (j_{1}\frac{1}{2} 10 |
j_{2}\frac{1}{2}) 
\end{eqnarray}
Thus $<p_{\frac{3}{2}} ||M1|| p_{\frac{1}{2}}> =
-\frac{1}{\sqrt{\pi}}(g_{s}-g_{\ell})$.  Here the bare values of the
parameters are:  $g_{\ell\pi} = 1,  g_{\ell\nu} = 0,  g_{s\pi} =
5.586, g_{s\nu} = -3.826$.

For particle-hole matrix elements of a tensor operator the same
reference [8] gives:
\begin{equation}
<[j_{1}^{-1}j_{2}]^{J} T_{\lambda}0> = (-1)^{j_{1}+j_{2}-\lambda}
<j_{2}||T_{\lambda}j_{1}> \delta_{J\lambda}
\end{equation}
In the spherical limit the $J = 1^{+}, T = 1$ state in $^{12}C$ is
\begin{equation}
\psi = \frac{[p_{\frac{1}{2}} p_{\frac{3}{2}}^{-1}]^{J=1}_{\pi} +
[p_{\frac{1}{2}} p_{\frac{3}{2}}^{-1}]^{J=1}_{\nu}}{\sqrt{2}}
\end{equation}
Our result is then
\begin{eqnarray}
B(M1) & = & \frac{1}{(2\pi)} [(g_{s\pi} - g_{s\nu}) - (g_{\ell \pi} -
g_{\ell\nu})]^{2} \nonumber \\
& = & 11.262 \mu_{N}^{2}
\end{eqnarray}

In the rotational model of Bohr and Mottelson the following result is
given for an M1 transition from a J=0, K=0 state
\begin{equation}
B(M1) = |(J=1^{+}K| M1 |J=0 K=0)|^{2} (2 - \delta_{K,o})
\end{equation}
the factor of 2 for K not equal to zero comes from the fact that one
can excite both K and -K.

For an intermediate deformation the levels of the previous section
become:
\begin{tabbing}
\= xxxxx \= xxxxxxxxxxxxxx \= xxxxxxxxxxxxxxxxxxxx \= xxxxxxxxxxxxxxx \kill

\> 2) \> N = 1 \> \ \ $Y_{11}\uparrow$ \> occupied \\
\> 3) \> N = 1 \> \ \ $cY_{10}\uparrow$ + $dY_{11}\downarrow$ \> occupied \\
\> 4) \> N = 1 \> $-dY_{10}\uparrow$ + $cY_{11}\downarrow$ \> empty \\
\end{tabbing}
In the asymptotic oblate limit $c=0, d=1$.  In the spherical limit $c =
\sqrt{\frac{2}{3}}, \ d = \sqrt{\frac{1}{3}}$.

One gets both K=1 and K=0 contributions to M1
\begin{eqnarray}
K & = & 1 \ \ a) \ \ \ <4|M1|\bar{3}> = -cdg^{\prime}_{s} + (c^{2}-d^{2})
g^{\prime}_{L} \sqrt{2} \nonumber \\ 
K & = & 1 \ \ b) \ \ \ <\bar{4}|M1|\bar{2}> = cg^{\prime}_{s} - \sqrt{2}
dg^{\prime}_{L} \nonumber \\ 
K & = & 0 \ \ c) \ \ \ <4|M1|3> = -cd(g^{\prime}_{s} - g_{L}^{\prime})
\nonumber \\ 
K & = & 0 \ \ d) \ \ \ <\bar{4}|M1|\bar{3}> = cd(g^{\prime}_{s} -
g^{\prime}_{L}) \nonumber \\ \nonumber
\end{eqnarray}
Here 
\begin{eqnarray}
g^{\prime}_{s} = \frac{(g_{s\pi}-g_{s\nu})}{\sqrt{2}} \ \ \ \
g^{\prime}_{L} = \frac{g_{\ell\pi}-g_{\ell\nu}}{\sqrt{2}} \nonumber
\end{eqnarray}

To get the total B(M1) we add up the sum of the squares of a) b) c)
and d).

\begin{eqnarray}
B(M1) & = & \frac{3}{4} \pi \{[-cd g^{\prime}_{s} + (c^{2}-d^{2})
g^{\prime}_{L} \sqrt{2}]^{2}  \nonumber \\
& + & [cg^{\prime}_{s} - \sqrt{2} dg^{\prime}_{L}]^{2} +
2c^{2}d^{2}(g^{\prime}_{s}-g^{\prime}_{\ell})^{2}\}
\end{eqnarray}
In the above expression the first two terms
come from excitations to K=1 states and the last to K=0 states.  This
becomes identical to the spherical expression previously given for $c
= \sqrt{\frac{2}{3}}, d\sqrt{\frac{1}{3}}$.

In the asymptotic oblate limit this becomes
\begin{equation}
B(M1) = \frac{3}{2\pi} (g_{L\pi} - g_{L\nu})^{2} = 0.4775 \mu^{2}_{N}
\end{equation}
Only the \underline{orbital} part of M1 contributes in this limit.
This is known as a scissors mode excitation.

We also consider an intermediate case with $\delta = -0.3$ where we use the
wavefunctions of Max Irvine [9].  For these c=0.25457 and d = 0.96706
we obtain B(M1) = 2.655 $\mu^{2}_{N}$ which is close to the
experimental value $B(M1) = 2.61 \mu^{2}_{N}$.

We now consider the operator relevant to the neutrino capture $\nu_{\mu} +
^{12}C \rightarrow \ ^{12}N + \mu^{-}$.  For the M1 allowed we consider
the operator $2st_{+}$.  We can get the result for this from the
previous section by dropping the facrtor $\frac{3}{4}\pi$ and by
setting $g^{\prime}_{s}$ to 2 and $g^{\prime}_{\ell}$ to zero.  Hence $B(S=1,
L=0) = 12c^{2}d^{2}+4c^{2}$.  In the asymptotic oblate limit c=0 so this
vanishes.  In the spherical limit this becomes $\frac{16}{3}$.  For an
intermediate deformation, $\delta = -0.3$ the value is 0.987 (we do
not include a coupling constant $C_{GT} = 1.251$ in our definition of
B). 
\medskip

\noindent 4)  The operators $rY_{1,k}t_{+},
r[Y_{1}\vec{s}]^{\lambda}_{k}t_{+}$ and
$r[Y_{1}\vec{\ell}]^{\lambda}_{k}t_{+}$
\medskip 

At finite momentum transfer we can get other contributions to the
neutrino cross section.  We here consider the operators listed in the
title.  The operator $rY_{1,k}t_{+}$
arises from the ``Forbidden Fermi'', $r[Y_{1}\vec{s}]^{\lambda}_{k}t_{+}$
from ``forbidden Gamow-Teller''.  The dipole orbital operator is
included for completeness.

We give results for the total strength for these operators in Table 1.
We will give less details than in the last section.  The following
formulae are useful for constructing the table.

a)  Expansion of the Dipole Spin Operator.
\bigskip

$[Y^{1}s]^{o}_{o} = -1/\sqrt{3} Y^{1}_{o}s_{o} + 1/\sqrt{6}
Y^{1}_{1}s_{-} - 1/\sqrt{6} Y^{1}_{-1}s_{+}$ \\

$[Y^{1}s]^{1}_{o} = \frac{1}{2} Y^{1}_{1}s_{-} + \frac{1}{2}
Y^{1}_{-1} s_{+}$ \\

$[Y^{1}s]^{2}_{o} = \sqrt{\frac{2}{3}} Y^{1}_{o}s_{o} +
\frac{1}{\sqrt{12}} Y^{1}_{1}s_{-} - \frac{1}{\sqrt{12}}
Y^{1}_{-1} s_{+}$ \\

$\sqrt{2} [Y^{1}s]^{1}_{1} = Y^{1}_{1}s_{o} +
\frac{1}{\sqrt{2}} Y^{1}_{o} s_{+}$ \\

$\sqrt{2} [Y^{1}s]^{2}_{1} = Y^{1}_{1}s_{o} - \frac{1}{\sqrt{2}}
Y^{1}_{o} s_{+}$ \\

$\sqrt{2} [Y^{1}s]^{2}_{1} = -Y^{1}_{1} s_{+}$ \\

Here
\begin{tabbing}
xxxxxxxxxxxxxxxxxxxxxxxxx \= xxxxxxxxxxxxxxxxxxxxxxxxxx \kill

$<\uparrow s_{o} \uparrow > = \frac{1}{2}$ \> $<\downarrow s_{o}
\downarrow > = - \frac{1}{2}$ \\
$<\uparrow s_{+} \downarrow > = 1$ \> $<\downarrow s_{-} \uparrow > = 1$
\\
\end{tabbing}
\bigskip

b)  The following radial integral is useful

\begin{equation}
\int Y^{\ell_{1}}_{m_{1}} Y^{\ell_{2}}_{m_{2}}Y^{\ell_{3}}_{m_{3}}
d\omega =
[\frac{(2\ell_{1}+1)(2\ell_{2}+1)(2\ell_{3}+1)}{4\pi}]^{\frac{1}{2}} \
(^{\ell_{1}\ell_{2}\ell_{3}}_{o \ o \ o}) \ 
(^{\ell_{1}\ell_{2}\ell_{3}}_{m_{1}m_{2}m_{3}})
\end{equation}
where the last two factors are 3j symbols.  The following radial
integrals are needed
\bigskip

$\int os \ r \ op \ r^{2}dr = \sqrt{\frac{3}{2}} b$

$\int op \ r \ od \ r^{2}dr = \sqrt{\frac{5}{2}} b$

$\int op \ r \ 1s \ r^{2}dr = b$ \bigskip \\
where harmonic oscillator wave functions have been used and $b =
\sqrt{\hbar/m\omega}$.

As in the previous section we consider transition to states with
quantum numbers J and K.  We form all particle-hole states

$K_{p}, K_{H} \ (K = K_{P}-K_{H}) \ K \geq 0$

We then calculate the matrix element $M(K_{P},K_{H},J)_{K}$.  For
given K, J we get the total strength S(J,K) such that 

\begin{equation}
\nonumber
S(J,K) = \sum_{K_{P}K_{H}} |M(K_{P},K_{H},J)|^{2}
\end{equation}

Note that we consider not only excitations from 0p to 1s ,0d but
also from 0s to 0p.  To show the results in as simple a way as
possible we multiply the strengths (i.e. squared matrix elements) by
$4\pi m\omega/\hbar$.

The results, which are shown in Table 1 are very simple and very
striking.  For the operators $rY^{1}t$, \ \ $r[Y^{1}s]^{\lambda}$ and
$r[Y^{1}\ell]^{\lambda}$ the total summed strengths are exactly the
same in the spherical limit and the asymptotic limit.  This is quite
different from the operator $st$, the strength of which is large in
the spherical limit but vanishes in the asymptotic limit.  Also for
the operator $\ell \tau$ (scissors mode) the strength in the
asymptotic limit is different then in the spherical case -- it is a
factor of three larger.

Note that for a given $\lambda$ in the spin-dipole case
$[rY^{1}s]^{\lambda}$ the summed strength is not the same in the two
limits.  In the asymptotic case the partial sums for $\lambda = 0, 1$
and $2$ are proportional to ($2\lambda + 1$) - 2.25, 6.75 and 11.25.
In the spherical case they are 3.25, 
8.25 and 8.75 respectively.  But the total sum is the same in the two
cases. 

The redistribution of the dipole strength (spin independent and
dependent) and energy shifts arising from deformation may lead to some
modifications in the inclusive neutrino cross-sections.  This is due
to the fact that there is energy dependence in the neutrino fluxes
used in the experiment and also due to the energy dependence of the
neutrino-nucleus interaction.  The basic conclusion that the dipole
strength is 
not affected by deformation is a very strong indication that the
deformation will not change the inclusive cross-section a lot while at
the same time the exclusive cross-sections to the $^{12}N$ g.s. will
be reduced considerably by the deformation.

The Cal-Tech group [5] did not include deformation effects explicitly
but simulated the consequences of these decreasing the ``allowed'' GT
strength in the neutrino raction by a factor of five without
altering the total dipole strength.  Our calculation justifies this in
part but there are residual effects of deformation on the dipole
strengths in the sense that the strengths to \underline{indivdual} J
values do depend on deformation.  Since the various J components are
split in energy this could affect the neutrino cross section.

Note that in the dipole-orbital case there is no contribution for
$\lambda=0$.  The operator for this case can be written as $C
\vec{r}\cdot \vec{\ell}$ where C is a constant.  We can write
$\vec{r}\cdot \vec{\ell} = \vec{r} \cdot \vec{r} \times \vec{p} =
(\vec{r} \times \vec{r}) \cdot \vec{p}$.  This vanishes because
$\vec{r} \times \vec{r} = 0$.  Alternatively we can say that $\vec{r}
\cdot \vec{\ell}$ given us the component of angular momentum in the
radial direction -- this is zero.

Note that in the case $[rY^{1}\vec{\ell}]^{\lambda=2}t$ one excites the
twist mode.  Like the other cases we get the same results for this
mode in the spherical and asymptotic limits.
\bigskip
\medskip

\noindent 5)  Proof of the Independence of the Summed Strengths for the
Operators $rY^{1}, r[Y^{1}\vec{s}]^{\lambda}$ and
$[rY^{1}\ell]^{\vec{\lambda}}$ on Deformation.
\medskip

We consider the most difficult case the spin-dipole.  We consider
transition from a $k=\frac{1}{2}$ orbit $xY^{1}_{o}\uparrow +
yY^{1}_{1}\downarrow$.  The summer strength is

\begin{equation}
S = \sum_{LMM_{s}\lambda K} |<LMM_{s}[Y^{1}s]^{\lambda}_{K}
xY^{1}_{o}\uparrow + yY^{1}_{1}\downarrow>|^{2}
\end{equation}
We expand $[Y^{1}s]^{\lambda}_{K} = \sum_{\alpha\beta}(11\alpha \beta
| \lambda K) Y^{1}_{\alpha} s_{\beta}$.

We use the orthonormality property of the Clebsh-Gordan coefficients
\begin{displaymath}
\nonumber
\sum_{\lambda k} (11 \alpha \beta |\lambda k)(11
\alpha^{\prime}|\beta^{\prime}\lambda k) = \delta_{\alpha\alpha^{1}}
\delta_{\beta\beta^{\prime}}
\end{displaymath}
to obtain

\begin{displaymath}
\nonumber
S = \sum_{LMM_{s\alpha\beta}} |<LMM_{s} Y^{1}_{\alpha} s_{\beta}
xY^{1}_{o}\uparrow + y Y^{1}_{1}\downarrow >|^{2}
\end{displaymath}

We can evaluate the spin part using $<\uparrow s_{o}\uparrow> =
\frac{1}{2} \ \ 
<\downarrow s_{o}\downarrow > = -\frac{1}{2} \ \ <\uparrow s_{1}
\downarrow > = -\frac{1}{\sqrt{2}} \ \  <\downarrow s_{1} \uparrow > =
\frac{1}{\sqrt{2}}$

\begin{eqnarray}
S & = & S(M_{s} = \frac{1}{2} , \beta = 0) + S(M_{s} = \frac{1}{2} ,
\beta = 1) \nonumber \\
& + & S(M_{s} = -\frac{1}{2} , \beta = 0) + S(M_{s} =
-\frac{1}{2} , \beta = -1) \nonumber \\
S & = & \sum_{\alpha LM} \{\frac{x^{2}}{4} |<LM Y^{1}_{\alpha}
Y^{1}_{o}>|^{2} + \frac{y^{2}}{2} |<LM Y^{1}_{\alpha} Y^{1}_{1}>|^{2}
\nonumber \\
& + & \frac{y^{2}}{4} |<LM Y^{1}_{\alpha} Y^{1}_{1} >|^{2} +
\frac{x^{2}}{2} |<LM Y^{1}_{\alpha} Y^{1}_{o} >|^{2}\} \nonumber \\ \nonumber
\end{eqnarray}
 
By using the Wigner-Eckart theorem we can show
\begin{eqnarray}
\sum_{M\alpha} |<LM Y^{1}_{\alpha} Y^{1}_{b} >|^{2} = \sum_{M\alpha}
|<LM Y^{1}_{\alpha} Y_{1} >|^{2} = \frac{1}{3} |<L || Y^{1} ||
Y^{1}>|^{2} \nonumber
\end{eqnarray}
Hence the summed strength becomes
\begin{eqnarray}
\nonumber
S = \frac{1}{4} \sum_{L} <L|| Y^{1}|| Y^{1}> (x^{2}+y^{2}) \nonumber
\end{eqnarray}
However for the orbit $x \ Y^{1}_{o}\uparrow + yY^{1}_{1}\downarrow$ \ the
normalization condition is $x^{2} + y^{2} = 1$.  Hense S is
independent of x and y and hence of deformation.

Note that the independence of deformation holds also separately for
each L.  The proofs for the other modes $rY^{1}$ and
$r[Y^{1}\ell]^{\lambda}$ go along the same lines as above and are in
fact much simpler.
\medskip

\noindent 6)  Summed Strength - An Alternate Approach
\medskip

We now show that the summed strength is independent of the shell model
configuration as long as all valence particles in the ground state are
in the 0p shell and that the expectation value of r.r is the same for
all particles in the p shell.

Let the dipole operator be O.  The summed strength is the ground state
expectation value of Sum(i,j)O(i).O(j).  If i does not equal j we get
zero if all nucleons are in the 0p shell because of parity.  If i
equals j then if O=RY$^{1}$ we get basically r.r and this is the same for
all particles.  For the spin dipole we get for J=0 s.r s.r which
equals 3/4 r.r.  For J=1 we get $\vec{s}\cdot \vec{r} \ \vec{s} \cdot
\vec{r}$ which equals 1/2 r.r.  For J=2 
we get (ss)J=2.(rr)J=2.  But consider (ss)J=2 Ms=2.  This equals 1/2
$s_{+} s_{+}$.  This vanishes because we cannot raise the spin of a spin 1/2
particle twice.  The other components of (ss)J=2 will also vanish.
Thus the ground state expectation value is reduced to one body terms
of the form r.r which will be the same for $p_{3/2}$ and $p_{1/2}$ particles
provided harmonic oscillator wave functions are used.  To get model
dependent results we have to include admixtures from higher shells
and/or use non harmonic oscillator radial integrals.

To make the argument complete we must also take into account the fact
that there are particles in the 0s shell and one can have matrix
elements of O(i).O(j) between 0s and 0p.  This will have the structure
of an exchange single particle interaction of a 0p particle with a
closed 0s shell.  This single particle matrix element will be the same
for $p_{3/2}$ as for $p_{1/2}$ only a 2 body spin-orbit interaction
can lead to 
a single particle splitting of $p_{3/2}$ and $p_{1/2}$.

Extensive experimental work on $^{12}C(n,p) ^{12}B$ reaction has been
performed by N. Olsson et. al [10].  Recently a preprint has appeared in
the LANL network by P. Vogel [11] in which detailed calculations show
that the total strength does not depend on $p_{1/2}$ shell occupancy.

This work was supported by the U.S. Department of Energy under grant
DE-FG02-95ER-40940.

\newpage
\centerline{Table 1 \ \ \  Isovector dipole strength in
$^{12}C(\frac{4\pi m\omega}{\hbar} B(\lambda)0^{+} \rightarrow \lambda)$}

\begin{tabbing}
xxxxxxxxxxxxxxxx\= xxxxxxxxxxxxxxxxxxx\= xxxxxxxxxxxxx\= xxxxxxx \kill

\underline{Dipole} \> \underline{$rY^{1}_{K}t$} \> \> \\
\> Asymptotic \> Spherical \>  \\
$\lambda$ \ \ K \> \> \> \\
0 \ \  0 \> \ \ \ \ \ \ \ \ 9 \> \ \ \ \ \ 9 \> \ \ \  \\
1 \ \  1 \> \ \ \ \ \underline{\ \ \ 18 \ \ \ } \>  \ \ \underline{
\ 18 \ \ \ } \> \bigskip \\ 
Sum \> \ \ \ \ \ \ \ 27 \> \ \ \ \ 27 \> \\
\\
\\
\\

\underline{Spin Dipole} \> \underline{$r[Y^{1}s]^{\lambda}t$} \> \> \\
\ \ $\lambda$ \ \ \ \  K \> \> \> \\
\ \ 0 \ \ \ \ 0 \> \ \ \ \ 2.25 \> \ \ \ 3.25 \> \\
\ \ 1 \ \ \ \ 0 \> \ \ \ \ 2.25 \> \ \ \ 2.75 \> \\
\ \ 1 \ \ \ \ 1 \> \ \ \ \ 4.50 \> \ \ \ 5.50 \> \\
\ \ 2 \ \ \ \ 0 \> \ \ \ \ 2.25 \> \ \ \ 1.75 \> \\
\ \ 2 \ \ \ \ 1 \> \ \ \ \ 4.50 \> \ \ \ 3.50 \> \\
\ \ 2 \ \ \ \ 2 \> \ \ \underline{\ \ 4.50 \ \ } \> \ \underline{ \ 3.50
\ } \>
\bigskip \\ 
\ \ Sum  \> \ \ 20.25 \> \ 20.25 \> \\
\\
\\
\\
\underline{Orbital} \> \underline{Dipole $r[Y^{1}\ell]^{\lambda}t$} \> \> \\
$\lambda$ \ \ K \> \> \> \\
0 \ \ 0 \> \ \ \ \ \ \ 0 \> \ \ \ 0 \> \\
1 \ \ 0 \> \ \ \ \ \ \ 3 \> \ \ \ 6 \> \\
1 \ \ 1 \> \ \ \ \ \ 11 \> \ \ \ 8 \> \\
2 \ \ 0 \> \ \ \ \ \ \ 9 \> \ \ \ 6 \> \\
2 \ \ 1 \> \ \ \ \ \ 19 \> \ \ 16 \> \\
2 \ \ 2 \> \ \ \ \ \underline{ \ \ 6 \ \ } \> \underline{\ \ 12 \ \ }
\> \bigskip \\ 
Sum \> \ \ \ \ \ 48 \> \ \ 48 \> \\

\end{tabbing}

\newpage
\centerline{Table 2 \ \ Same as Table 1 but only for 0s $\rightarrow$ 0p
Transitions}

\begin{tabbing}
xxxxxxxx\= xxxxxxxxxxxxxxxx \= xxxxxxxxxxxxxxxxxxxxx \=
xxxxxxxxxxxxxxxxxxxxxx \kill 

\> \> Asymptotic \> Spherical \\
\\
\> $\lambda$ \ \ K \> $[rY^{1}S]^{\lambda}_{K}$ \> \\
\> 0 \ \ 0 \> \ \ \ 0.25 \> \ \ \ 0.75 \\
\> 1 \ \ 0 \> \ \ \ \ 0 \> \ \ \ 0.50 \\
\> 1 \ \ 1 \> \ \ \ 0.75 \> \ \ \ 1.00 \\
\> 2 \ \ 0 \> \ \ \ 0.5 \> \ \ \ 0.00 \\
\> 2 \ \ 1 \> \ \ \ 0.75 \> \ \ \ 0.00 \\
\> 2 \ \ 2 \> \underline{ \ \ 0.00 \ \ } \> \underline{ \ \ 0.00 \ \ }
\bigskip \\
\> \> \ \ \ 2.25 \> \ \ \ 2.25 \\
\\
\\
\\
\> \> \ \ \ $rY^{1}$ \> \\
\> $\lambda$ \ \ K \> \> \\
\> 1 \ \ 0 \> \ \ \ \ 3 \> \ \ \ \ 1 \\
\> 1 \ \ 1 \> \ \underline{ \ \ 0 \ \ }\> \ \underline{ \ \ 2 \ \ } \\
\> \> \ \ \ \ 3 \> \ \ \ \ 3 \\
\end{tabbing} 

\newpage
\centerline{References}
\bigskip

\begin{enumerate}
\item J. N. Bahcall, Neutrino Astrophysics (Cambridge University
Press, New York, 1989), J. N. Bahcall and M. H. Pinsonneault,
Rev. Mod. Phys. \underline{64}, 885 (1992).

\item C. Athanassopoulos et al., Phys. Rev. Lett. \underline{77},
3082 (1976).

\item B. Zeitmitz, KARMEN Collaboration, Progr. Part. Nucl. Phys,
\underline{32}, 351 (1994).

\item C. Athanassopoulos et al., Phys. Rev. C\underline{56}, 2806
(1997); W. C. Louis, private communication.

\item E. Kolbe, K. Langanke, F. K. Thieleman, and P. Vogel,
Phys. Rev. C\underline{52}, 3437 (1995).

\item N. Auerbach, N. V. Giai, and D. K. Vorov,
Phys. Rev. C\underline{56}, R2368 (1997).

\item S. Cohen and D. Kurath, Nucl. Phys. \underline{73} (1965) 1.

\item A. Bohr and B. Mottelson, Nuclear Structure, Vol. 1 and Vol. 2
(1969) and (1975), (Benjamin, New York, 1975).

\item J. M. Irving, Nuclear STructure Theory, Oxford Pergamon Press
(1972). 

\item N. Olsson et. al., Nucl. Phys. A\underline{559}, 368 (1993).

\item P. Vogel, Neutrino Nucleus Scattering, LANL nucl-th/9901027,
1999. 
\end{enumerate}

\end{document}